# Proactive AI-and-RAN Workload Orchestration in O-RAN Architectures for 6G Networks

Syed Danial Ali Shah[1,2] Maryam Hafeez[1] Abdelaziz Salama[1] Syed Ali Raza Zaidi[1]
[1]School of Electronic and Electrical Engineering, University of Leeds, Leeds, UK
[2]Adelaide University/University of South Australia

***Abstract*—The vision of AI-RAN convergence, as advocated by the AI-RAN Alliance, aims to unlock a unified 6G platform capable of seamlessly supporting AI and RAN workloads over shared infrastructure. However, the architectural framework and intelligent resource orchestration strategies necessary to realize this vision remain largely unexplored. In this paper, we propose a Converged AI-and-ORAN Architectural (CAORA) framework based on O-RAN specifications, enabling the dynamic coexistence of real-time RAN and computationally intensive AI workloads. We design custom xApps within the Near-Real-Time RAN Intelligent Controller (NRT-RIC) to monitor RAN KPIs and expose radio analytics to an End-to-End (E2E) orchestrator via the recently introduced Y1 interface. The orchestrator incorporates workload forecasting and anomaly detection modules, augmenting a Soft Actor-Critic (SAC) reinforcement learning agent that proactively manages resource allocation, including Multi-Instance GPU (MIG) partitioning. Using real-world 5G traffic traces from Barcelona, our trace-driven simulations demonstrate that CAORA achieves near 99% fulfillment of RAN demands, supports dynamic AI workloads, and maximizes infrastructure utilization even under highly dynamic conditions. Our results reveal that predictive orchestration significantly improves system adaptability, resource efficiency, and service continuity, offering a viable blueprint for future AI-and-RAN converged 6G systems.**

***Index Terms*—O-RAN, 6G, AI, xApp, RIC, SAC.**

## I. Introduction

THE rapid expansion of mobile communications and the growing demand for network capacity are driving innovations in the next-generation wireless network architecture. To improve resource utilization and harness maximum value from the scarce spectral resource, increased virtualisation, disaggregation, and densification have become core ingredients of network design. While there are several alternative architectural frameworks, one possible architectural choice that has democratised the research for Next-G wireless networks is the Open Radio Access Network (O-RAN) architecture. O-RAN has transformed traditional cellular networks by enabling open interfaces and multi-vendor interoperability, paving the way for efficient resource utilization and dynamic network optimization [1], [2]. However, these advancements also introduce new challenges as networks evolve to support increasingly diverse workloads. In particular, with the growing deployment of AI & ML applications alongside traditional RAN workloads, there is a pressing need for intelligent resource management systems that can effectively share computing infrastructure between these diverse workload types [3]–[5]. Current RAN infrastructures often maintain dedicated resources for network functions, leading to under-utilization during off-peak periods while simultaneously struggling to accommodate the growing demands of AI workloads.

AI and RAN, one of the three key domains envisioned by the AI-RAN Alliance, aims to support the coexistence of AI and RAN workloads on shared RAN infrastructure [4], [6]. This involves deploying AI applications on RAN infrastructure, leveraging shared accelerated compute, e.g., GPU and memory resources, to run intensive AI tasks while simultaneously supporting RAN operations [4]. The coexistence of RAN and AI workloads on shared infrastructure presents unique challenges due to their distinct characteristics, dynamic nature, and lifecycle management requirements. Cellular networks are inherently dynamic, where RAN workloads continuously vary, making real-time tracking of resource demands essential for optimal resource allocation, in particular for the co-existence scenarios [7], [8]. In contrast, AI workloads are computationally intensive and exhibit variable resource demands depending on their specific use cases and deployment contexts.

Traditional static or reactive resource allocation mechanisms are inadequate for the coexistence of dynamic AI and RAN workloads. In highly variable wireless environments, where sudden traffic surges and fluctuating compute demands are common, a purely reactive approach can lead to resource contention, service degradation, and underutilization of critical infrastructure. Therefore, a proactive resource management framework capable of forecasting workload patterns and anticipating demand variations is essential to ensure optimal resource allocation, maintain service continuity, and maximize infrastructure efficiency.

Some recent advances and AI-based solutions have shown promise in dynamic resource management scenarios. However, existing solutions focus either on RAN optimization or AI workload scheduling independently and are insufficient for AI and RAN coexistence scenarios. The authors in [9] presented a comprehensive power consumption model for O-RAN configurations to understand the energy consumption patterns in O-RAN architecture. In [10], authors presented Radio Intelligent Controller (RIC) modules integrating meta-learning for real-time network data collection and dynamic management of RAN resources. An O-RAN-compatible adversarial learning-based resource allocation scheme is proposed in [11] to enhance the energy efficiency of virtualised base stations and O-RAN components. Some other relevant works introduce advanced O-RAN testbeds and frameworks, e.g., Open AI Cellular (OAIC) framework for designing and testing AI-based RAN Management Algorithms [12], and X5G, a private 5G O-RAN Testbed with NVIDIA ARC and OpenAirInterface [13],

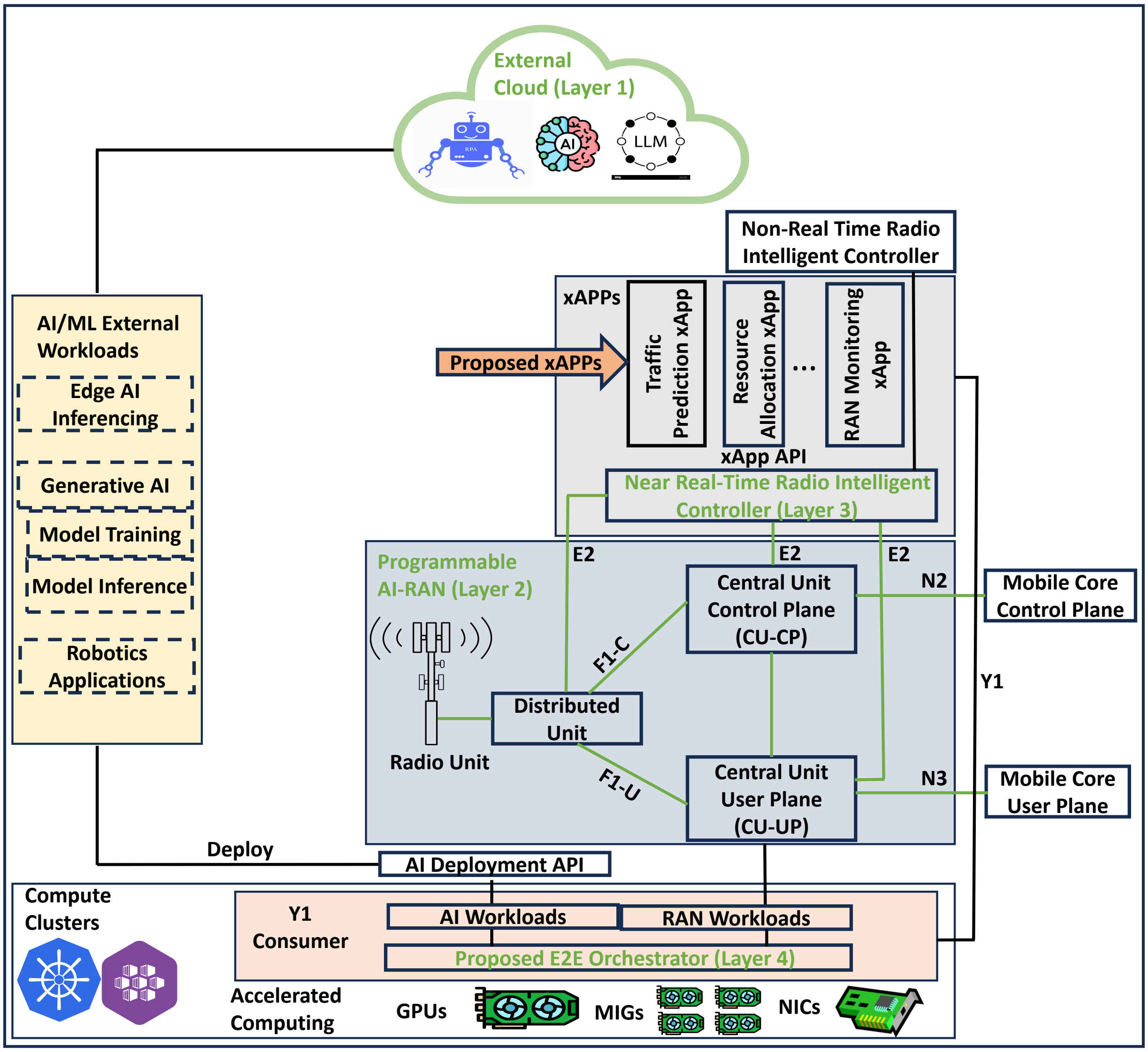

Fig. 1. The proposed converged AI-and-ORAN architectural framework

[14].

However, none of the above state-of-the-art works and frameworks address the architectural framework or optimization requirements necessary to support the coexistence of AI and RAN on shared infrastructure. Furthermore, the dynamic nature of cellular networks, the real-time demands of RAN network functions, and the varying computational needs of AI workloads necessitate a more sophisticated approach capable of proactively and dynamically adjusting resource allocation policies based on the changing requirements.

This paper presents a novel architectural framework based on O-RAN specifications to address the requirements for dynamic, predictive resource allocation and the coexistence of RAN and AI workloads on shared RAN infrastructure. The key working principles of the proposed framework and its contributions are summarized as follows:

- **Converged AI-and-ORAN Architectural Framework:** To explore the convergence of both compute-and-communication, we propose a Converged AI-and-ORAN Architectural (CAORA) framework based on O-RAN specifications, incorporating custom-built xApps within the Near-Real-Time RAN Intelligent Controller (NRT-RIC). This framework advances the evolving AI and RAN domain by enabling the coexistence of AI and RAN workloads on shared RAN infrastructure. The xApp continuously tracks RAN KPIs and exposes the radio analytics information with the proposed End-to-End (E2E) orchestrator through the recently introduced Y1 interface, enabling dynamic data collection and network monitoring for intelligent resource management decisions.
- **Intelligent E2E Orchestrator and Cross-Layer Resource Management Framework:** We introduced a system that establishes communication between xApps integrated within the NRT-RIC and the external E2E orchestrator via the Y1 interface. This framework enables real-time adaptation of resource allocation policies based on evolving network dynamics. The proposed E2E orchestrator employs a Soft Actor-Critic (SAC) reinforce-ment learning algorithm, acting as a Y1 consumer to communicate with the NRT-RIC. It processes real-time metrics from the monitoring xAPP, maintains historical network patterns, and implements dynamic resource allocation policies, thereby enabling the coexistence of RAN and AI workloads on shared RAN infrastructure.

- **Enhanced Proactive Orchestration:** We introduce *SpikeAwareLSTM*, a multi-task Long Short-Term Memory (LSTM) architecture that jointly performs time-series forecasting and anomaly detection, e.g., traffic spikes, enabling a forecast-aware SAC agent to proactively and dynamically allocate shared resources. This anticipatory orchestration significantly improves adaptability and infrastructure utilization in AI-RAN coexistence scenarios.
- **Trace-Driven Evaluation Under Realistic and Diverse Conditions:** We validate our proposed architecture using real-world traces from Barcelona City's 5G network, which include highly dynamic scenarios such as sudden demand spikes during football matches (Les Corts – Camp Nou), steady traffic from residential zones (Poble Sec), and fluctuating loads in touristic nightlife areas (El Born). This diverse dataset enables rigorous testing of the framework's adaptability and robustness under varying and unpredictable workload patterns. Through simulations, we demonstrate improved utilization of shared RAN infrastructure and consistent RAN performance under dynamic workloads. Compared to baseline schemes, our approach better supports AI–RAN coexistence and achieves superior adaptability and proactive resource allocation under varying demand conditions.

The remainder of this paper is organized as follows: Section II reviews the related works; Section III introduces the proposed converged architecture; Section IV provides a detailed system model and problem formulation; Section V discusses the predictive resource demand model based on LSTM; Section VI describes the dataset used for evaluation; Section VII outlines the simulation environment and presents the performance evaluation; and, finally, Section VIII concludes the paper.

## II. Related Works

In recent years, substantial interest has been shown in integrating AI techniques within the O-RAN framework. Polese *et al.* [15] presented ColO-RAN, a flexible testbed enabling the deployment of Deep Reinforcement Learning (DRL)-based xApps for real-time slicing and scheduling optimization in O-RAN environments. Zhang *et al.* [16] proposed a federated DRL approach for collaborative resource management across distributed RICs, enabling privacy-preserving yet efficient network slicing. Lotfi *et al.* [17] combined traffic prediction with distributed DRL to improve slice admission control and resource utilization under dynamic demand patterns. Firouzi and Rahmani [18] introduced a hierarchical reinforcement learning framework for managing resource allocation in IoT-oriented network slices, showing improved convergence compared to flat RL models. Kouchaki *et al.* [19] demonstrated practical deployment strategies for RL-based xApps on NRT-RICs, focusing on standardized interfaces and data pipelines. Similarly, Barker *et al.* [20] developed an O-RAN Gym-based system to streamline the training and validation of DRL models on O-RAN platforms.

Beyond single-agent models, Attanayaka *et al.* [21] proposed a peer-to-peer federated learning mechanism for detecting anomalies without centralized coordination, improving resilience in distributed RICs. Bakri *et al.* [22] introduced a graph convolutional network-based multi-agent reinforcement learning (MARL) approach for scalable coordination among multiple xApps. Rumesh *et al.* [23] further embedded a hierarchical federated learning structure within a digital twin of O-RAN to enable more efficient model updates across RIC tiers. More recently, Large Language Models (LLMs) have been investigated by Tahir *et al.* [24] and Wu *et al.* [25], showcasing how LLMs can enhance decision-making and resource management flexibility by incorporating semantic context into RIC policies.

However, despite these advances, none of the above-mentioned works have focused on the coexistence of AI-driven tasks and conventional RAN workloads on the same shared RAN infrastructure. In particular, the dynamic reallocation of computational resources, e.g., GPU cores, between RAN and AI workloads based on real-time network states remains largely unexplored. Addressing this gap is critical to enabling more efficient and intelligent O-RAN systems that maximize infrastructure utilization without degrading RAN performance.

## III. Converged AI-and-ORAN Architectural Framework

We present the CAORA framework to enable the coexistence of RAN and AI workloads on shared RAN infrastructures, as shown in Fig. 1. CAORA implements a hierarchical control structure comprising four key layers: The external cloud, programmable RAN components, the controller, and resource management layers. The external cloud layer (layer 1) hosts AI/ML workloads, e.g., edge AI inferencing, generative AI, and robotics applications, that interface with the RAN infrastructure, i.e., compute clusters, through the AI Deployment API. The programmable RAN layer (layer 2) includes disaggregated RAN components based on the O-RAN architecture, e.g., Distributed Units (O-DU), O-RAN Radio Units (O-RU), and O-RAN Central Units (O-CU). The controller layer (layer 3) consists of the NRT-RIC, which integrates with the proposed xApps for traffic prediction, resource allocation, and network monitoring, facilitating rapid resource allocation decisions via the E2 interface. The proposed xApps, e.g., monitoring xApp, are also responsible for consistently tracking and monitoring the network load information to determine the RAN workload requirements and communicating it with the proposed E2E orchestrator.

The E2E Orchestrator implements a SAC reinforcement learning algorithm interfacing with the monitoring xApp, deployed within the NRT-RIC through the Y1 interface, as shown in Fig. 1. The Y1 interface connects the NRT-RIC with an authorized Y1 consumer for exposure to radio analytics information. The monitoring xApp continuously collects critical network metrics, e.g., bandwidth utilization, system latency, and network load, and feeds this data into the SAC agent, i.e., the Y1 consumer. The SAC agent uses this information to implement the priority-based and dynamic allocation of shared computing resources, where RAN workloads receive precedence during peak periods while AI workloads utilise available resources during off-peak times, supporting the coexistence of the RAN and AI workloads on shared RAN infrastructure, as illustrated in Fig. 1. The proposed SAC agent supports the coexistence of RAN and AI workloads on shared infrastructure through three key mechanisms: predictive analysis utilizing historical patterns for resource planning, real-time adaptation adjusting allocations based on current conditions as informed by the monitoring xApp, and priority management ensuring RAN performance while maximizing AI workload support.

The resource management layer (layer 4) implements the proposed E2E orchestrator that manages the RAN infrastructure resources, i.e., compute clusters containing GPUs, Multi-Instance GPUs (MIGs) based on NVIDIA's MIG technology [26], and Network Interface Cards (NICs). The proposed architectural framework transforms traditional RAN infrastructure into a multi-purpose system that efficiently supports both network operations and AI workloads.

TABLE I
SUMMARY OF KEY NOTATIONS

| Symbol | Description |
|---|---|
| $K$ | Set of all tasks (RAN/AI workloads) |
| $\tau$ | Task type ($\tau \in \{\text{RAN}, \text{AI}\}$) |
| $\boldsymbol{r}(t)$ | Instantaneous resource demand vector |
| $p$ | Task priority level |
| $\hat{\boldsymbol{r}}_{\text{LSTM}}(t+\delta)$ | LSTM-predicted demand for $\delta$-step lookahead |
| $\mathcal{F}_\theta$ | LSTM prediction model with parameters $\theta$ |
| $\theta$, $\phi$ | LSTM and SAC policy parameters |
| $T_{\text{hist}}$ | Historical window length for LSTM inputs |
| $\delta$, $H$ | Prediction horizon index and max. horizon |
| $\mathcal{R}_{\max}$ | Maximum system resource capacity |
| $d_x(t)$ | Current demand for service $x \in \{\text{RAN}, \text{AI}\}$ |
| $\hat{d}_x(t+\delta)$ | Predicted future demand for service $x$ at $t+\delta$ |
| $r_x(t)$ | Resource allocation for service $x$ at time $t$ |
| $\nu_x^{\max}$ | Max. allocation adjustment rate for service $x$ |
| $\Delta r_x$ | Allocation adjustment action for service $x$ |
| $\gamma$ | Discount factor for future rewards |
| $\alpha_\delta$ | Borrowing coefficient for future horizon $\delta$ |
| $\beta$ | Future contention discount factor |
| $\eta$ | Contention penalty coefficient |
| $R'_t$ | Proactive reward function at time $t$ |
| $\mu$ | Anticipation weight in reward function |
| $\lambda$ | Overprovisioning penalty coefficient |
| $\kappa$ | Non-linear penalty curvature |
| $C_x(t)$ | Completed resources for service $x$ at $t$ |
| $P_c(k,t)$ | Task completion probability |
| $C_{\text{pred}}(t)$ | Predictive contention factor |
| $s_t$ | Binary spike indicator |
| $\tau_{\text{spike}}$ | Threshold for spike detection in demand |
| $\lambda_{\text{detect}}$ | Spike detection loss weighting coefficient |
| $\hat{s}_t^{\text{int}}$ | Internal spike probability from SpikeAwareLSTM |
| $\hat{\Delta}_{\text{free}}(t+\delta)$ | Predicted free resources at $t+\delta$ |

## IV. SYSTEM MODEL WITH PREDICTIVE ALLOCATION

### A. AI/RAN Task Representation with LSTM Forecasting

Let $K$ denote the set of all tasks in the system, including RAN and AI workloads. Each task $k \in K$ is characterized by:

$$k = \{\tau, \boldsymbol{r}, p, \hat{\boldsymbol{r}}_{\text{LSTM}}\}, \tag{1}$$

where $\tau \in \{\text{RAN}, \text{AI}\}$ is task type that identifies whether an individual task belongs to the RAN or AI workload category, $r(t) \in \mathbb{R}_+^N$ represents the instantaneous resource demands, e.g., GPU, $p \in [0,1]$ is the priority level representing the real-time criticality of the tasks, and $\hat{\boldsymbol{r}}_{\text{LSTM}}(t+\delta)$ is the LSTM-predicted demands for $\delta$-step lookahead.

The tasks are categorized into two service groups: $x = \text{RAN}$, which represents real-time RAN workloads, e.g., signal processing and resource allocation, and $x = \text{AI}$, which refers to compute-intensive AI workloads, e.g., distributed inference and generative AI. While each individual task is labeled by a task-type indicator $\tau \in \{\text{RAN}, \text{AI}\}$, we use the variable $x \in \{\text{RAN}, \text{AI}\}$ more generally to express equations and constraints at the service group level. This abstraction allows us to define aggregate resource demands and priorities for all tasks of a specific type in a unified way. Service group priorities $p_x$ for $x \in \{\text{RAN}, \text{AI}\}$ are derived as:

$$p_x(t) = \max_{k \in K_x(t)} p(k), \tag{2}$$

where $K_x(t)$ is the set of tasks of type $x$ active at time $t$.

The LSTM prediction model is formalized as:

$$\hat{\boldsymbol{r}}_{\text{LSTM}}(k, t+\delta) = \mathcal{F}_\theta\left(\{\boldsymbol{r}(k,\tau)\}_{\tau=t-T_{\text{hist}}}^{t}\right). \tag{3}$$

The LSTM predictor $F_\theta$ is implemented as a `SpikeAwareLSTM`, which is discussed later in the paper. This model jointly optimizes demand forecasting and spike detection. During training, it learns to suppress spurious predictions in the presence of anomalous spikes, ensuring that $\hat{r}_{LSTM}$ inherently reflects these spike conditions.

### B. Predictive System Dynamics

The workload evolution (Eq. 4) models temporal resource dynamics by fusing real-time ($\|\boldsymbol{r}\|_1$) and LSTM-predicted ($\|\hat{\boldsymbol{r}}_{\text{LSTM}}\|_1$) demands via Bayesian averaging. The predictive contention factor $C_{\text{pred}}$ accounts for both current and discounted future resource contention ($\beta$-weighted), ensuring allocations adapt to anticipated workload fluctuations. This formulation captures the interplay between task generation, completion, and proactive resource arbitration in dynamically evolving AI and RAN systems. The workload evolution incorporates multi-step predictions:

$$W(t) = \int_0^t \left(\mathbb{E}[\Delta(k,\tau)|\mathcal{F}_\theta] - P_c(k,\tau)\Delta(k,\tau)C_{\text{pred}}(k,\tau)\right) d\tau, \tag{4}$$

where, $\mathbb{E}[\Delta|\mathcal{F}_\theta] = \frac{1}{2}\left(\|\boldsymbol{r}\|_1 + \|\hat{\boldsymbol{r}}_{\text{LSTM}}\|_1 \cdot (1 - \hat{s}_t^{\text{int}})\right)$ represents the Bayesian demand fusion that integrates the spike detection factor. Specifically, $\hat{s}_t^{\text{int}} \in [0,1]$ denotes the internal spike probability from the `SpikeAwareLSTM`, which dynamically adjusts the reliance on predictions during detected anomalies. This internal value, though not exposed to the SAC agent, plays a critical role in workload modeling.

Additionally, $C_{\text{pred}}(k,t) = \frac{\Delta_{\text{total}}(t) + \sum_{\delta=1}^{H} \beta^\delta \hat{\Delta}_{\text{total}}(t+\delta)}{\mathcal{R}_{\max}}$ defines the predictive contention factor, where $\beta \in (0,1)$ is the future contention discount factor (e.g., $\beta = 0.9$).

To proactively allocate resources, we predict the availability of free resources based on the expected completion of ongoing tasks. Let $\hat{\Delta}_{\text{free}}(t+\delta)$ represent the total resources anticipated to be freed at time $t+\delta$, computed from LSTM-based predictions of task lifetimes and resource demands:

$$\hat{\Delta}_{\text{free}}(t+\delta) = \sum_{k\in K} P_c(k,t+\delta)\Delta(k,t+\delta)r(k,t+\delta), \quad (5)$$

where $P_c(k,t+\delta) \in [0,1]$ is the completion probability of task $k$ by $t+\delta$, predicted by LSTM or based on historical data, $\Delta(k,t+\delta) \in \mathbb{R}_+$ denotes the resources allocated to task $k$ at $t+\delta$, based on the scheduling policy, and $r(k,t+\delta) \in \mathbb{R}_+^N$ represents the resource demand rate, e.g., GPU usage, of task $k$ at $t+\delta$, predicted by LSTM. The product $P_c \cdot \Delta \cdot r$ estimates the expected freed resources for task $k$, considering both its completion likelihood and resource requirements. Summing over all tasks $k \in K$ provides the predicted system-wide available resources for future allocation.

### C. Predictive-Adaptive Optimization Problem

The objective (Eq. 6) maximizes the *contention-aware expected cumulative reward* $\mathbb{E}\left[\sum_{t=0}^{T}\gamma^t\left(R'_t - \eta C_{\text{pred}}(t)\right)\right]$, balancing immediate service quality (via $R'_t$) with long-term demand anticipation and resource contention penalties. Constraint (7) enables predictive resource reservation by augmenting the physical resource limit $\mathcal{R}_{\max}$ with anticipated freed resources $\sum_{\delta=1}^{H}\alpha_\delta\hat{\Delta}_{\text{free}}(t+\delta)$, where $\hat{\Delta}_{\text{free}}$ represents LSTM-predicted resources released from completed tasks. This ensures proactive allocation without violating hardware limits. Constraint (8) enforces rate-limited allocation adjustments via $\|r_x(t)-r_x(t-1)\| \le v_x^{\max}\Delta t$, ensuring smooth transitions between allocation states to avoid destabilizing RAN/AI workloads.

$$\max_{r_{\text{RAN}},r_{\text{AI}}} \mathbb{E}\left[\sum_{t=0}^{T}\gamma^t\left(R'_t - \eta C_{\text{pred}}(t)\right)\right] \quad (6)$$

$$\text{s.t.} \quad r_{\text{RAN}} + r_{\text{AI}} \le \mathcal{R}_{\max} + \sum_{\delta=1}^{H}\alpha_\delta\hat{\Delta}_{\text{free}}(t+\delta) \quad (7)$$

$$\|r_x(t) - r_x(t-1)\| \le v_x^{\max}\Delta t \quad \forall x \in \{\text{RAN},\text{AI}\} \quad (8)$$

### D. LSTM-Enhanced SAC Architecture

*1) Predictive State Space:* The state $s_t$ integrates both *current* and *predicted* RAN/AI demands ($d_x(t)$, $\{\hat{d}_x(t+\delta)\}$) to enable proactive resource allocation. The inclusion of prior allocations $r_x(t-1)$ ensures temporal consistency, allowing the SAC agent to learn from historical decisions. This holistic representation of system dynamics balances real-time requirements with anticipatory planning, empowering the agent to optimize allocations across finite and forecasted resource budgets. The state space is given as:

$$s_t = \left(d_{\text{RAN}}(t), d_{\text{AI}}(t), \{\hat{d}_x(t+\delta)\}_{\delta=1}^{H}, r_x(t-1)\right), \quad (9)$$

where $d_x(t) = \sum_{k\in K_x}\|\boldsymbol{r}(k,t)\|_1$ represents the current resource demand, $\hat{d}_x(t+\delta)$ is the future resource demand predicted by the LSTM ($\delta = 1,\ldots,H$), and $r_x(t-1)$ is the previous allocation, ensuring temporal consistency.

*2) Proactive Constrained Reward Function:* The proposed reward function balances the current service quality, measured by the completed-to-demanded resource ratios, with the proactive anticipation of future workloads using LSTM predictions. A non-linear penalty term is introduced to enforce resource capacity constraints, exponentially penalizing both current and predicted allocations that exceed the system's capacity. This design encourages the SAC agent to optimize immediate performance while preparing for forecasted demand spikes, ensuring that hardware safety margins are maintained. The parameters $(\mu,\beta,\kappa)$ govern the tradeoff between reactivity and foresight.

The reward function combines two key components: the current QoS, $\left(\frac{C_x}{d_x}\right)$, and the anticipated fulfillment of predicted demands, $\left(\frac{\mu\cdot\min(p_x r_x,\hat{d}_x)}{d_x}\right)$, as predicted by the LSTM model. To manage overprovisioning, a penalty term applies a super-linear cost (controlled by $\kappa = 2.5$) for resource allocations $R_{\text{alloc}}(t)$ that exceed the normalized capacity $\mathcal{R}_{\max}$. The reward function is given by:

$$R'_t = \underbrace{\sum_{x\in X}\left(\frac{C_x(t)}{d_x(t)} + \mu\frac{\hat{C}_x(t+1)}{d_x(t)}\right)}_{\text{Current + Anticipated QoS}} - \lambda\left[\left(\frac{R_{\text{alloc}}(t)+\beta\hat{R}_{\text{alloc}}(t+1)}{\mathcal{R}_{\max}}\right)^{\kappa} - 1\right] - \eta C_{\text{pred}}(t). \quad (10)$$

The first term in the reward function reflects the combined effect of current and anticipated QoS. The second term penalizes overprovisioning by applying a super-linear cost to resource allocations exceeding normalized capacity. The parameters $\mu$, $\beta$, and $\lambda$ control the balance between anticipation, future discounting, and the penalty for overprovisioning. The contention penalty $\eta C_{\text{pred}}(t)$ uses the predictive contention factor from Eq. 4, ensuring allocations avoid periods of high current or forecasted resource contention.

The pseudocode for the proposed CAORA framework is provided in Algorithm 1, while the different phases involved in the framework are illustrated in Figure 2.

## V. SpikeAwareLSTM: A Multi-Task Learning Approach for Time-Series Forecasting and Anomaly Detection

In this work, we propose a multi-task learning approach for time-series forecasting, which integrates anomaly detection (spike detection) directly within the LSTM architecture. This dual-task framework addresses the challenges of predicting temporal sequences while simultaneously identifying rare but significant anomalies often indicative of critical system changes, e.g., sudden peaks in network load, thereby offering a holistic solution to real-time data analysis. Traditional time-series models and anomaly detection techniques are typically treated as separate tasks; however, we leverage the inherent temporal dependencies captured by LSTMs to jointly optimize both prediction accuracy and spike detection. We introduce a composite loss function that combines Mean Squared Error (MSE) for time-series prediction with Binary Cross-Entropy (BCE) for spike detection, encapsulated in the following equation:

$$\begin{aligned} L =& \frac{1}{T}\sum_{t=1}^{T}\left(\boldsymbol{r}(k,t) - \hat{\boldsymbol{r}}_{\text{LSTM}}(k,t)\right)^2 \\ &+ \lambda_{\text{detect}} \cdot \frac{1}{T}\sum_{t=1}^{T}\Big[s_t \log(\hat{s}_t^{\text{int}}) \\ &+ (1-s_t)\log(1-\hat{s}_t^{\text{int}})\Big]. \end{aligned} \tag{11}$$

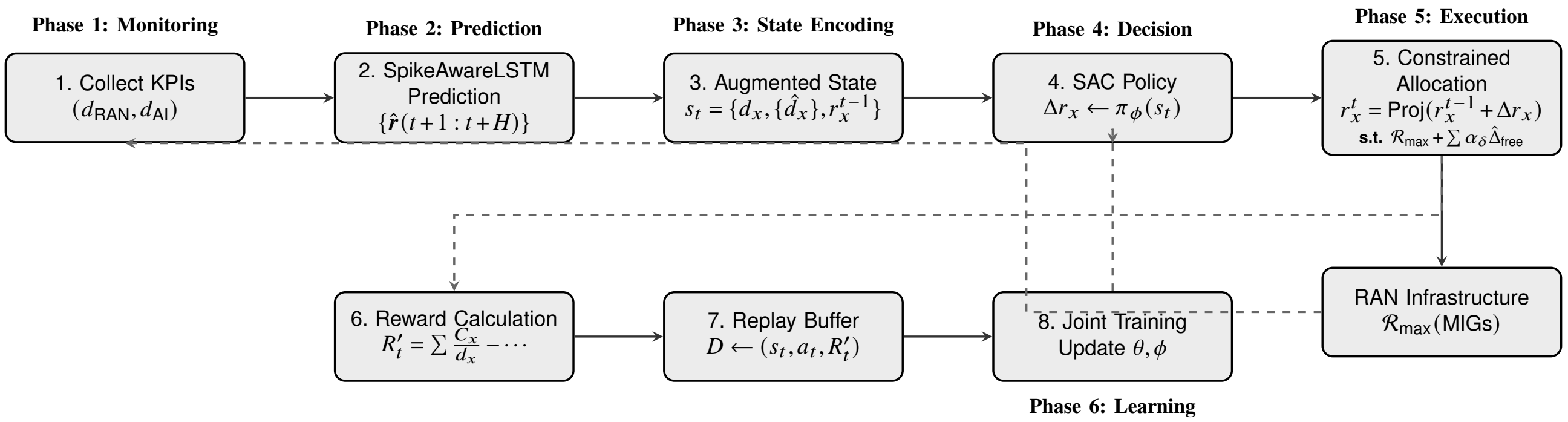

Fig. 2. LSTM-SAC resource orchestration framework for AI-and-RAN systems

The loss function $L$ integrates two objectives: resource demand prediction and spike detection. The first term focuses on resource demand prediction, where the MSE is calculated between the true resource demand vector $\boldsymbol{r}(k,t)$ (defined in Eq. 1) and the LSTM-predicted resource demand $\hat{\boldsymbol{r}}_{\text{LSTM}}(k,t)$ generated by $\mathcal{F}_\theta$ (Eq. 3). This term encourages the model to minimize the difference between the true and predicted resource demands, ensuring accurate forecasting over time. The second term in the loss function is concerned with spike detection and uses BCE. The binary spike indicator $s_t \in \{0,1\}$ is derived from historical demand data, where $s_t = 1$ if $\|\boldsymbol{r}(k,t)\|_1 > \tau_{\text{spike}}$ for a predefined threshold $\tau_{\text{spike}}$, and 0 otherwise. By treating spike detection as a binary classification problem, the model is trained to classify each data point as either a spike or not, with a dedicated output layer employing a sigmoid activation to map the LSTM's internal states to spike probabilities. The spike detection output is given by:

$$\hat{s}_t^{\text{int}} = \sigma(W_z h_t + b_z),$$

where $\sigma$ denotes the sigmoid function, $W_z$ is the weight matrix for the spike detection layer, $h_t$ is the LSTM's hidden state at time $t$, and $b_z$ is the bias term. The LSTM architecture's ability to retain memory over long sequences is pivotal in detecting non-linear, rapid changes indicative of spikes. The model adapts dynamically to historical data trends, enabling it to recognize both underlying patterns and sudden deviations, thereby enhancing its ability to predict and flag high-magnitude spikes that exceed normal fluctuations. The predicted internal spike probability $\hat{s}_t^{\text{int}} \in [0,1]$ is generated by the LSTM model, reflecting its confidence in whether a spike occurs at time $t$. This probability dynamically adjusts the reliance on LSTM predictions during anomalies (see Eq. 4), ensuring robust resource arbitration under volatile workloads. The BCE term penalizes incorrect spike predictions by comparing the actual spike indicator $s_t$ with the predicted spike probability $\hat{s}_t^{\text{int}}$. The hyperparameter $\lambda_{\text{detect}}$ is used to control the trade-off between the importance of accurate resource demand prediction and spike detection.

### A. Complexity Analysis

The computational complexity of the LSTM-SAC orchestration framework is governed by two core components: the LSTM predictor and the SAC reinforcement learning agent. The LSTM model, structured with two layers ($64 \rightarrow 32$ units), introduces a time complexity of $O(T_{\text{hist}} \cdot (U_1^2 + U_2^2 + U_2))$ per timestep, where $T_{\text{hist}}$ is the historical window length, and $U_1 = 64$ and $U_2 = 32$ are the number of LSTM units. The additional $O(T_{\text{hist}} \cdot U_2)$ term accounts for spike detection at each timestep, with the associated memory overhead increasing marginally by approximately 1 KB due to the spike detection parameters $(W_z, b_z)$.

For the SAC agent, the three-layer MLP-based actor and twin critic networks (with 128 hidden units per layer) contribute a complexity of $O(B \cdot (|\mathcal{S}| \cdot 128 + 128^2 + 128 \cdot |\mathcal{A}|))$ per training iteration, where $B = 64$ is the batch size, $|\mathcal{S}|$ is the augmented state dimension, and $|\mathcal{A}| = 2$ is the action space size. The use of twin Q-networks doubles this cost.

Memory overhead arises from the replay buffer (which stores 100,000 transitions, requiring approximately 4 MB) and the storage of historical KPIs for LSTM inputs. While joint training of the LSTM and SAC introduces additive computational demands, the framework's rate-limited allocation adjustments ($v_x^{\max}$) and efficient MIG resource partitioning ensure real-time operational feasibility. These design choices enable scalable deployment in O-RAN environments, even under dynamically fluctuating RAN and AI workloads.

## VI. Dataset Preparation and Evaluation of Prediction Models

To ensure practical relevance, we integrate real-world datasets to drive workload dynamics and system evaluation. Specifically, we extract workload patterns and KPI traces, e.g., Radio Network Temporary Identifier (RNTI) count, from

**Algorithm 1** LSTM-SAC Proactive Resource Orchestration for AI-and-RAN Systems

**Input:** $d_{\text{RAN}}(t)$, $d_{\text{AI}}(t)$, $\mathcal{R}_{\max}$, $T_{\text{hist}}$, $H$
**Output:** Optimal $r_{\text{RAN}}(t)$, $r_{\text{AI}}(t)$
**Initialize:**
  NRT-RIC, xApp, E2E Orchestrator
  LSTM pred. $\mathcal{F}_\theta$, SAC agent $\pi_\phi$, Replay buffer $D$
**for** each time step $t$ **do**
  **1. Collect KPIs:**
  $d_{\text{RAN}}, d_{\text{AI}} \leftarrow$ xApp (latency, throughput, load)
  $p_x(t) \leftarrow \max_{k \in K_x} p(k)$ ▷ Priority aggregation
  **2. LSTM Prediction:**
  $\{\hat{\boldsymbol{r}}(t+1:t+H)\} \leftarrow \mathcal{F}_\theta(\boldsymbol{r}(t-T_{\text{hist}}:t))$
  $\hat{d}_x(t+\delta) = \sum_{k \in K_x} \|\hat{\boldsymbol{r}}(k, t+\delta)\|_1$ ▷ $\delta = 1, \dots, H$
  **3. Update Augmented State:**
  $s_t = \{d_{\text{RAN}}, d_{\text{AI}}, \{\hat{d}_x(t+\delta)\}, r_{\text{RAN}}(t-1), r_{\text{AI}}(t-1)\}$
  **4. SAC Policy Decision:**
  $a_t = \{\Delta r_{\text{RAN}}, \Delta r_{\text{AI}}\} \leftarrow \pi_\phi(s_t)$
  **5. Constrained Resource Allocation:**
  $\Delta r_x \leftarrow \text{clip}(a_t[\Delta r_x], -\nu_x^{\max}\Delta t, +\nu_x^{\max}\Delta t)$
  $r_x^{\text{temp}}(t) = \text{Proj}_{[0,\mathcal{R}_{\max}]}\left(r_x(t-1) + \Delta r_x\right)$
  $R_{\text{total}} = \mathcal{R}_{\max} + \sum_{\delta=1}^{H} \alpha_\delta \hat{\Delta}_{\text{free}}(t+\delta)$
  **if** $r_{\text{RAN}}^{\text{temp}} + r_{\text{AI}}^{\text{temp}} > R_{\text{total}}$:
   Scale $r_x(t) \propto p_x(t)$ to satisfy $\sum x r_x(t) \le R_{\text{total}}$
  **else**:
   $r_x(t) \leftarrow r_x^{\text{temp}}(t)$
  **6. Proactive Reward Calculation:**
  $\hat{R}_{\text{alloc}}(t+1) \leftarrow \sum_x \hat{d}_x(t+1)$
  $R'_t = \sum_x \left(\frac{p_x(t)C_x(t)}{d_x(t)} + \mu \frac{\min(p_x(t)r_x(t), \hat{d}_x(t+1))}{d_x(t)}\right)$
  $-\lambda\left[\left(\frac{R_{\text{alloc}}(t) + \beta\hat{R}_{\text{alloc}}(t+1)}{\mathcal{R}_{\max}}\right)^\kappa - 1\right] - \eta C_{\text{pred}}(t)$
  **7. Update Replay Buffer:**
  $D \leftarrow (s_t, a_t, R'_t, s_{t+1})$
  **8. Joint Training:**
  Sample mini-batch $(s, a, R', s') \sim D$
  Update Q-networks: $Q_i \leftarrow \arg\min \mathbb{E}[(Q_i(s,a) - (R' + \gamma \min Q_j(s', \pi(s'))))^2]$
  Update policy: $\nabla_\phi J(\pi) \approx \mathbb{E}[\nabla_\phi \pi(s) \nabla_a Q(s,a)]$
  Update LSTM: $\nabla_\theta \mathcal{L}_{\text{pred}} = \mathbb{E}[\|\hat{\boldsymbol{r}} - \boldsymbol{r}\|_2^2 + \lambda \cdot \text{BCE}(\hat{s}_t, s_t)]$
**end for**
**Return:** Optimal allocations $r_{\text{RAN}}(t)$, $r_{\text{AI}}(t)$

5G traffic datasets from the Barcelona city cellular trace. These data streams are preprocessed into time-series profiles representing temporal workload demands for different task classes. The trace-based KPIs are fed into the SAC state vector, allowing the learning agent to react to real network dynamics such as congestion and QoS violations. This integration ensures that our orchestration strategy is robust under realistic conditions. We considered data from various locations in Barcelona (Spain), which provide unique network utilization patterns influenced by both daily activities and special events. The following open-source datasets [27], [28] were used for our analysis:

- **Les Corts - Camp Nou (LCCM):** A residential area near Camp Nou, the stadium of FC Barcelona, which frequently hosts soccer matches and special events. The model was trained on measurements collected over a 12-day period (from January 12, 2019, at 17:12 to January 24, 2019, at 16:20), including three soccer matches. The evaluation was conducted using data from a separate day.
- **Poble Sec (PS):** A residential area situated between key landmarks, including the historic center, Montjuïc mountain, and the port. The model was trained on data collected over a 28-day period (from February 5, 2018, at 23:40 to March 5, 2018, at 15:16). The evaluation was conducted using data from a separate day.
- **El Born (EB):** A touristic area adjacent to the city's old center, characterized by its vibrant amusement and nightlife. The model was trained on data collected over a 7-day period (from March 28, 2018, at 15:56 to April 4, 2018, at 22:36). The evaluation was conducted using data from a separate day.

TABLE II
LSTM SIMULATION PARAMETERS

| Parameter Category | Configuration |
|---|---|
| **Model Architecture** | |
| - LSTM Structure | 2 layers (64 → 32 units) |
| - Activation | ReLU (LSTM), Linear/Sigmoid (outputs) |
| - Regularization | Dropout (0.2), LayerNorm (1st LSTM) |
| - Output Heads | RNTI count (regression), Spike detection (binary) |
| **Training Setup** | |
| - Sequence Length | 10 timesteps |
| - Batch Size | 256 samples |
| - Epochs | 1000 |
| - Optimization | Adam ($\alpha = 0.001$) |
| - Loss Functions | MSE (RNTI), BCE (spikes) |
| **Data Configuration** | |
| - Input Feature | RNTI count (normalized) |
| - Preprocessing | StandardScaler (Z-score) |
| - Split | Separate train/test CSV files |
| **Spike Detection** | |
| - Threshold | $90^{\text{th}}$ percentile |
| - Classification | Binary (1=spike, 0=normal) |

We trained our proposed model using the RNTI count, which serves as an indicator of demand for RAN resources. The simulation parameters used for training and testing our predictive model are summarized in Table II. Our model outperforms the state-of-the-art approach presented in the literature [28] in accurately predicting RAN demand, particularly in terms of detecting peak congestion levels. Accurate spike detection is crucial for ensuring efficient allocation of shared RAN infrastructure between RAN and AI workloads, especially in dynamic scenarios where congestion fluctuates.

As demonstrated in Fig. 3, Fig. 4, and Fig. 5, the proposed predictive model accurately forecasts both RAN demand and peak congestion levels. These predictions are essential for the proactive and dynamic allocation and distribution of resources between AI and RAN workloads in coexistence scenarios.

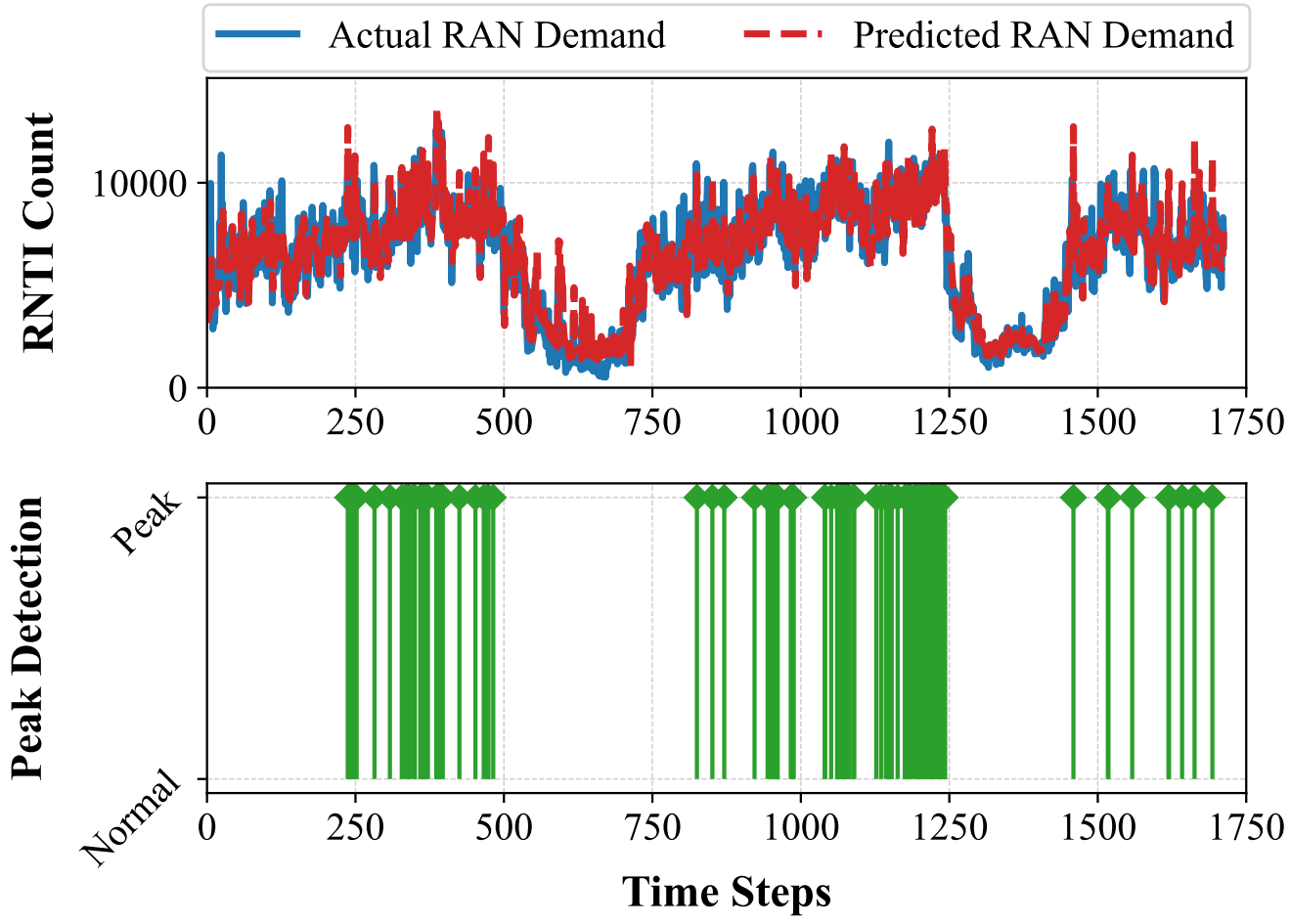

Fig. 3. Resource demand prediction for Les Corts-Camp Nou

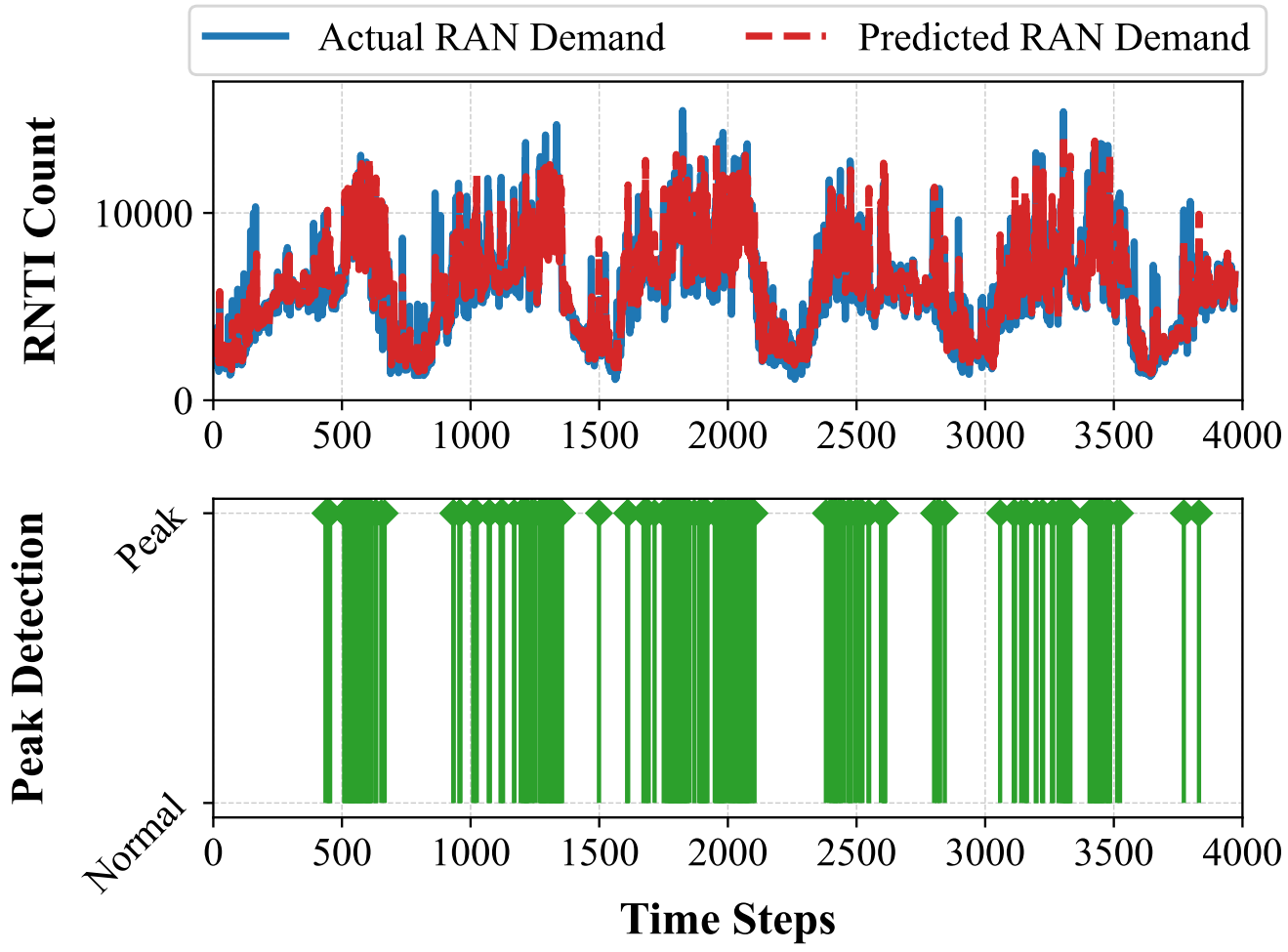

Fig. 4. Resource demand prediction for Poble Sec

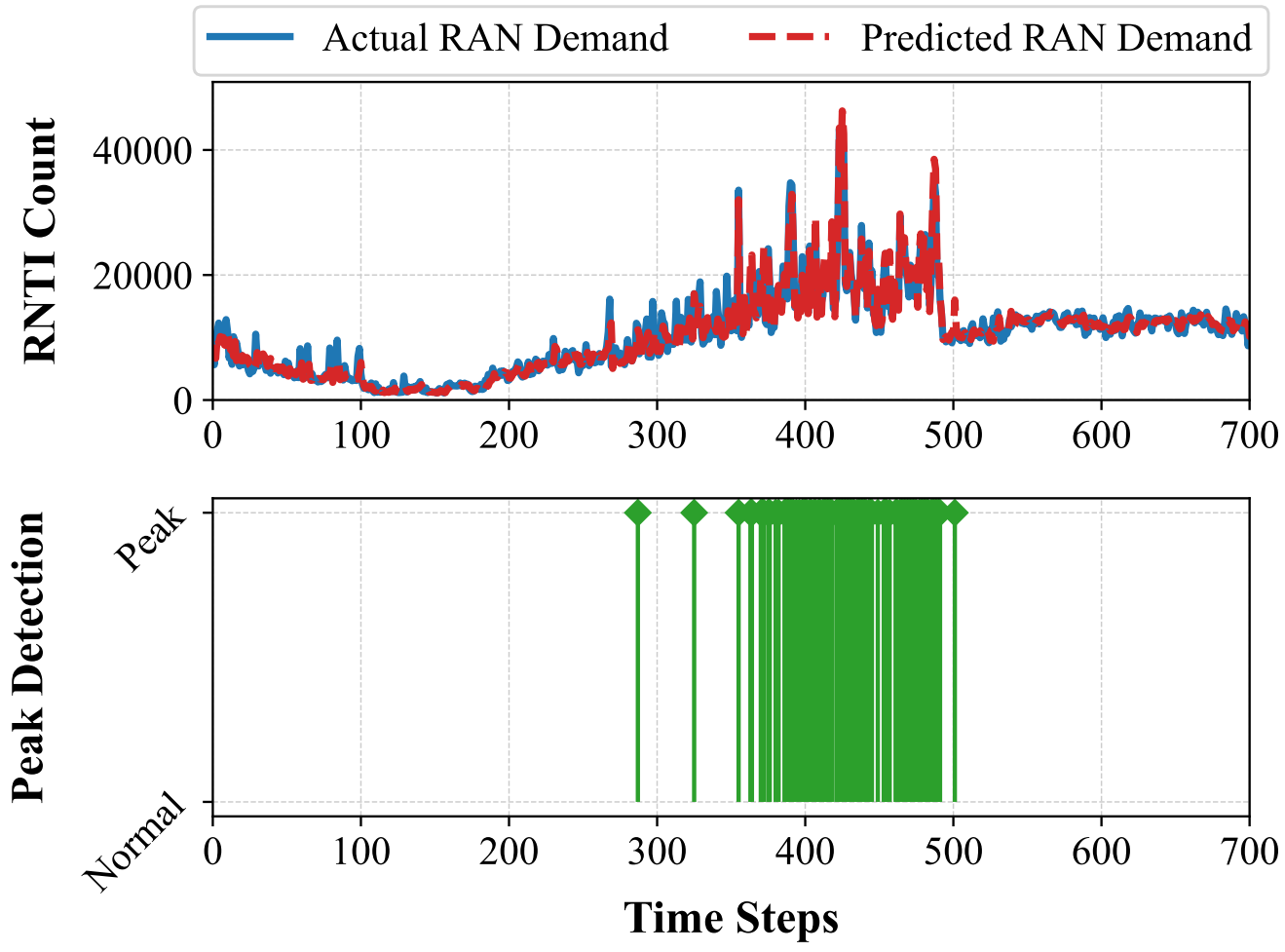

Fig. 5. Resource demand prediction for El Born

## VII. Performance Evaluation

### A. Experimental Setup and Implementation

The experimental evaluation of the proposed framework was conducted using Python, leveraging the O-RAN 7.2x split [1]. We simulated realistic O-RAN infrastructure dynamics, incorporating shared computing resources between RAN and AI workloads. The monitoring xApp, integrated as a Python module with the NRT-RIC, continuously tracks key network performance indicators, e.g., resource demands $d_{\text{RAN}}(t)$ from gNB-served micro-cells. This data is then fed into the proposed E2E orchestrator, i.e., layer 4 in the CAORA Framework. The orchestrator utilizes a two-layer LSTM predictor for $H$-step demand forecasting, while a SAC agent, using actor-critic networks, dynamically allocates resources. At each timestep $t$, the LSTM predicts future demands $\hat{\boldsymbol{r}}(t+1:t+H)$, which are incorporated into the SAC agent's state $s_t$. The SAC policy then generates allocation actions $a_t = \{\Delta r_{\text{RAN}}, \Delta r_{\text{AI}}, \nu_{\text{RAN}}, \nu_{\text{AI}}\}$, enforcing rate-limited projections and proactively optimizing rewards. The joint training of the LSTM and SAC enables adaptive resource partitioning between RAN and AI workloads. This integration facilitates real-time prioritization of RAN workloads during periods of congestion, while maximizing AI resource utilization during off-peak intervals, as demonstrated in the experimental results presented later.

In this paper, resources are modeled as Multi-Instance GPUs (MIGs) based on NVIDIA's MIG technology [26], where each GPU instance has dedicated resources allocated to specific workloads, such as AI or RAN tasks. NVIDIA's MIG technology allows partitioning a single GPU into isolated instances, each with dedicated memory, cache, and compute cores, ensuring efficient resource utilization [26]. For the simulation environment, we adopt the MIG 1g.5gb profile of the NVIDIA A100-SXM4-40GB GPU, which supports up to seven MIG instances, each with one-eighth of the total GPU memory allocated [26].

The resource allocation environment utilizes dual demand patterns to replicate realistic network conditions. RAN demand is derived from normalized RNTI counts, where temporal RNTI values are scaled linearly to the $[0,1]$ interval through min-max normalization. The scaled demand values $[0,1]$ are then interpreted as the proportion of maximum available resources required by the demand. The RAN demand is defined as:

$$d_{\text{RAN}}(t) = \frac{\text{RNTI}_t - \min(\text{RNTI})}{\max(\text{RNTI}) - \min(\text{RNTI}) + \epsilon}, \tag{12}$$

In contrast, AI demand follows a synthetic representative periodic pattern $D_t^{\text{AI}}$, designed to emulate the behavior of modern AI workloads such as large language model inference, batch processing, or scheduled retraining tasks [29]. These workloads typically exhibit cyclic compute intensity due to time-triggered execution or fluctuating user interaction patterns [29]. Accordingly, we model AI demand using a full-wave rectified sinusoidal function:

TABLE III
SIMULATION PARAMETERS AND MIG CONFIGURATION

| Parameter | Value |
|---|---|
| GPU Model | NVIDIA A100-SXM4-40GB |
| MIG Profile | 1g.5gb (7 instances) |
| Max Resources ($R_{\max}$) | 21 MIGs |
| Episode Length ($T$) | Dataset-dependent |
| Training Episodes | 1000 |
| Time Steps per Episode | 100 |
| Actor/Critic Learning Rate ($\alpha_{\text{actor}}/\alpha_{\text{critic}}$) | 3e-4 |
| Discount Factor ($\gamma$) | 0.99 |
| Target Update Rate ($\tau$) | 0.005 |
| Temperature Parameter ($\alpha$) | 0.2 |
| Replay Buffer Capacity | 100,000 |
| Hidden Layer Size | 128 |
| Training Batch Size | 64 |
| Actor/Critic Architecture | 3-layer MLP |

$$d_{\text{AI}}(t) = \frac{\sin\left(\frac{4\pi t}{N}\right) + 1}{2}, \tag{13}$$

where $N$ represents the total number of timesteps. This function yields two full oscillations over the evaluation horizon and is normalized to the $[0, 1]$ scale to ensure consistency with the RAN demand model. The chosen representation captures the periodic resource bursts commonly observed in GPU-driven AI tasks while maintaining simplicity and generality for simulation [29]. The contrasting nature of RAN's stochastic real-world variability and AI's structured, periodic behavior creates realistic resource contention dynamics within the system. The simulation parameters used throughout our evaluation are detailed in Table III.

## B. Performance Evaluation

*1) Evaluation Metrics:* We considered three key performance metrics: *completion rate*, *adaptability*, and *proactive resource allocation efficiency*, to assess the effectiveness of the proposed CAORA framework.

- **Completion rate**: Measures the percentage of demand successfully fulfilled by the system for both RAN and AI workloads. A higher completion rate indicates that the system is effectively matching available resources to task requirements in real time.
- **Adaptability**: Captures the system's ability to dynamically adjust resource allocations in response to shifting workload demands. In our context, adaptability reflects how closely the actual resource distribution between RAN and AI tasks follows the ideal demand ratio over time. A high adaptability score signifies efficient coordination between the system's decision-making agent and the observed workload variations. The adaptability metric is defined as:

$$\text{Adaptability} = 1 - \frac{1}{T}\sum_{t=1}^{T}\left|\alpha_t^* - \alpha_t\right|, \tag{14}$$

Here, $\alpha_t^*$ represents the ideal allocation ratio for RAN or AI workloads, while $\alpha_t$ denotes the actual allocation implemented. This metric assesses how closely the allocation strategy matches the demand-driven optimal values.
- **Proactive Resource Allocation Efficiency**: Assesses the system's proactive ability to allocate resources in advance, based on predictions of resource demand. This predictive approach helps minimize resource contention and improve service quality, especially during dynamic demand scenarios, e.g., special events, higher network load in tourist areas, or adapting to low-load periods in residential areas. It ensures efficient use of shared resources, effectively distributing them between AI and RAN workloads.

*2) Baseline Comparison:* As this work represents one of the first efforts to enable the coexistence of AI and RAN on a shared RAN infrastructure through dynamic orchestration, there is no directly comparable prior research. However, to benchmark our approach, we compare it against two baseline allocation strategies that do not incorporate dynamic learning or prediction.

- **Balanced Strategy**: This strategy evenly divides the available resources between AI and RAN workloads, allocating 50% to each, regardless of temporal demand variations. It represents a fairness-driven approach.
- **RAN Priority Strategy**: Reflects a priority-driven model, allocating 70% of the resources to RAN workloads and the remaining 30% to AI tasks. This prioritization is driven by the latency-sensitive nature of RAN services.

While both baseline strategies offer predictable behavior, they lack the ability to adapt to real-time traffic patterns or anticipated future demands. In contrast, our proposed method dynamically adjusts allocations based on observed and predicted load fluctuations, using a SAC reinforcement learning policy in conjunction with LSTM-based workload forecasting.

## C. Simulation Results

*1) Proactive Resource Allocation:* The efficacy of the proposed CAORA framework in dynamically allocating resources between RAN and AI workloads is validated through trace-driven simulations across three distinct scenarios: Les Corts - Camp Nou (LCCM), El Born (EB), and Poble Sec (PS). These scenarios, derived from real-world 5G traffic traces in Barcelona, each exhibit unique demand patterns, providing a rigorous stress test for the framework's ability to proactively adapt to dynamic and unpredictable workloads.

In the **Les Corts - Camp Nou (LCCM)** scenario, shown in Figure 6, resource allocation is tested during a football match at Camp Nou, where abrupt RAN demand spikes occur due to increased user density and real-time communication needs. The SpikeAwareLSTM model detects anomalies, such as RNTI surges, and provides predictive demand signals to the SAC agent in the proposed E2E orchestrator, i.e., layer 4 in the CAORA framework. The Y1 interface facilitates the monitoring xApp's communication of real-time KPIs to the E2E orchestrator, enabling anticipatory resource adjustments. During peak periods, the SAC agent prioritizes RAN workloads by allocating around 80% of MIG instances, while real-locating unused resources to AI tasks during quieter periods, as illustrated in Figure 6.

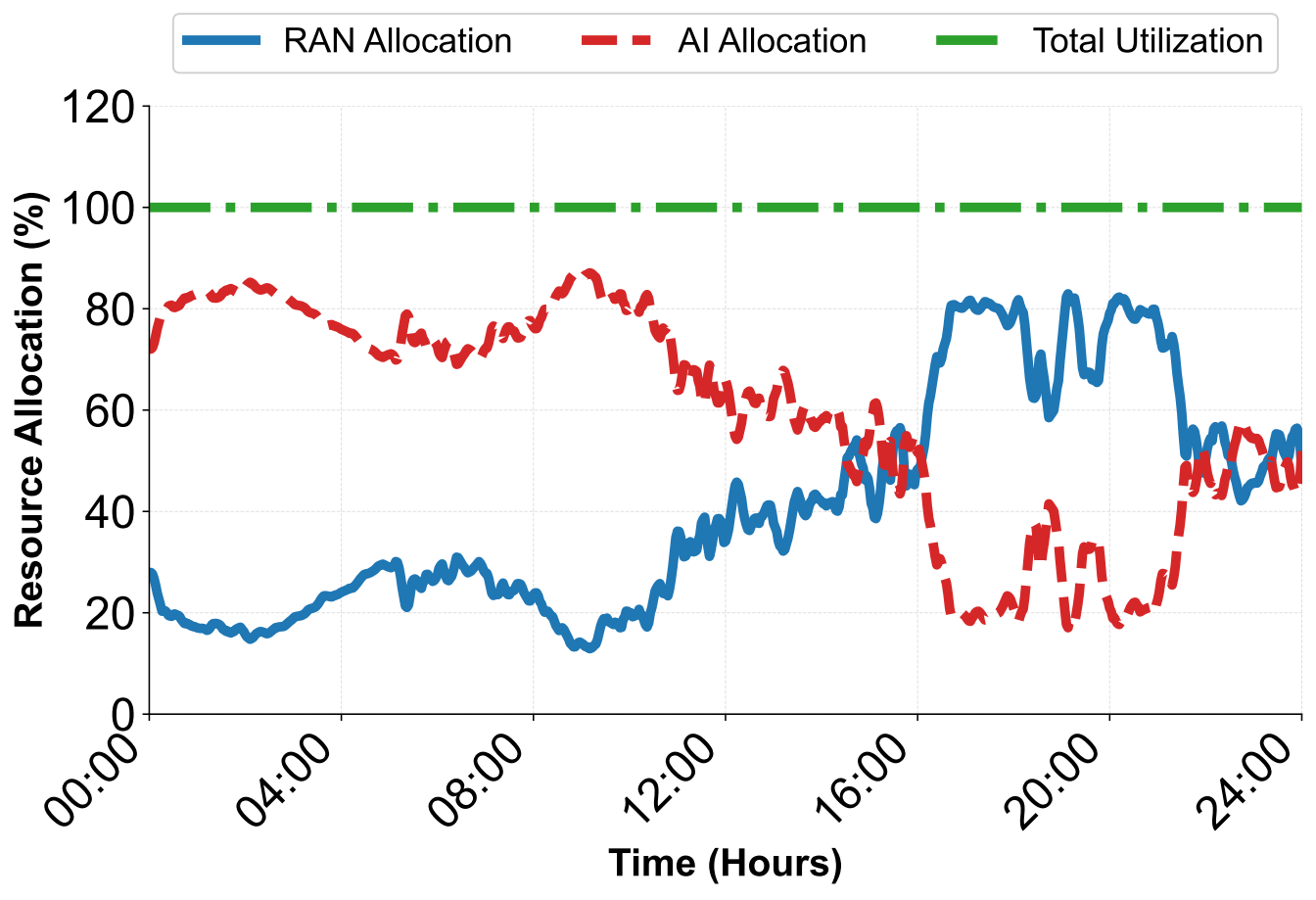

Fig. 6. Resource allocation during a football match (LCCM), prioritizing RAN to manage sudden demand spikes

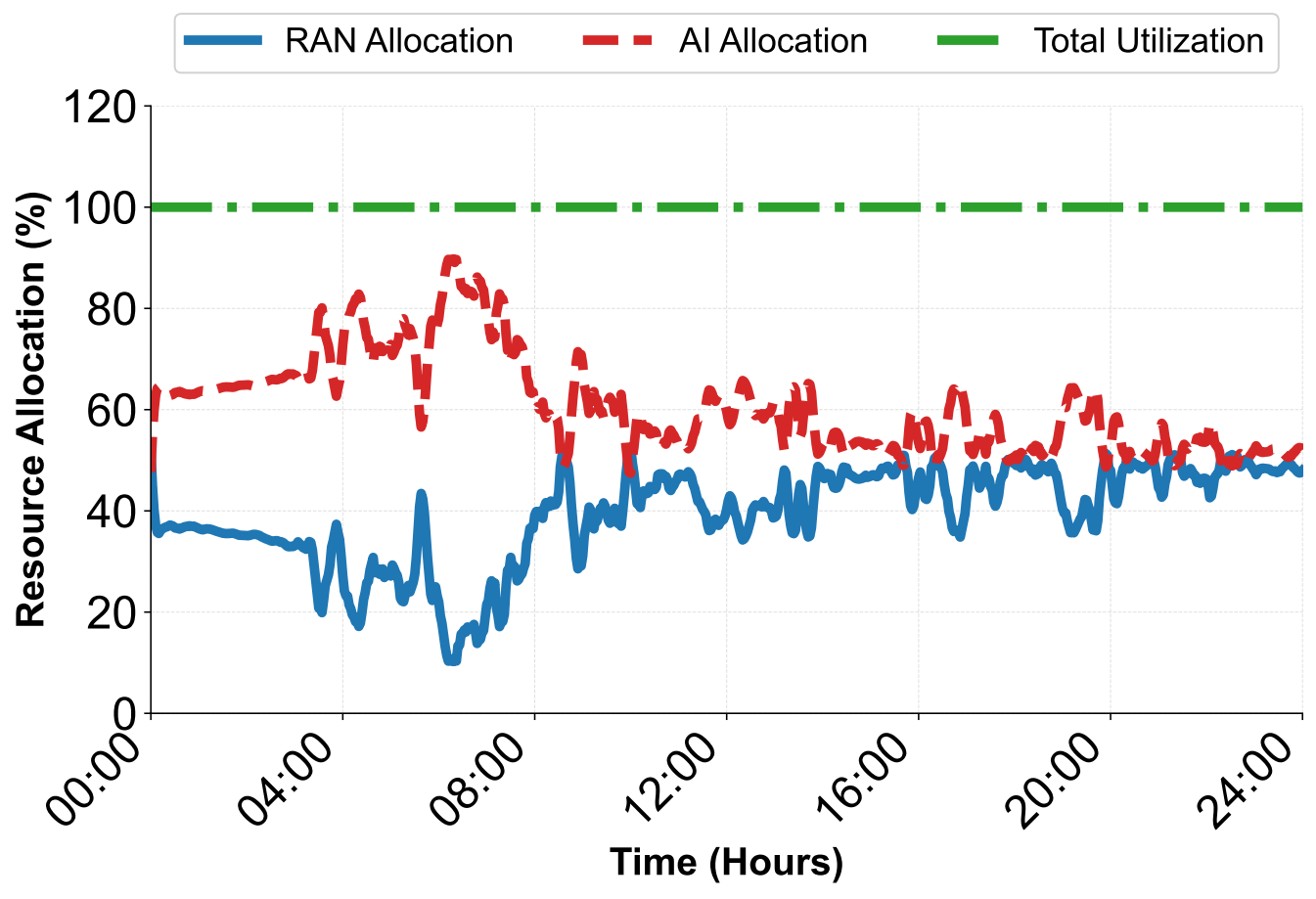

Fig. 8. Efficient allocation in a residential area (Poble Sec) with stable and low demand traffic patterns

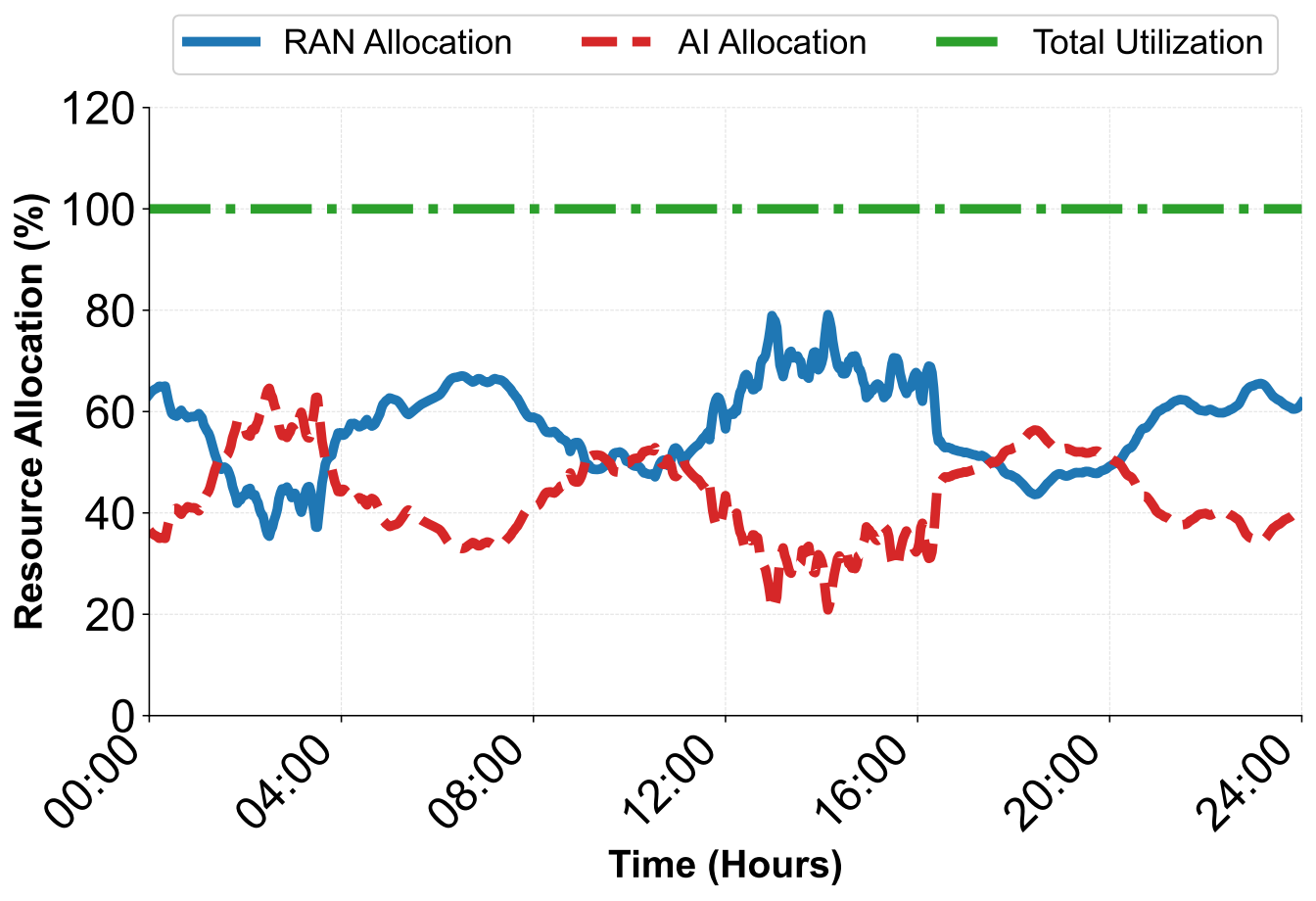

Fig. 7. Dynamic resource balancing for touristic and nightlife-driven demand fluctuations (El Born)

The **El Born (EB)** scenario, depicted in Figure 7, captures the cyclical demand fluctuations typical of the area's touristic and nightlife activities. These fluctuations lead to significant demand peaks in the evening, driving high RAN resource requirements. The LSTM predictor identifies these diurnal patterns, enabling the SAC agent to preemptively reserve resources for anticipated increases in RAN load. During off-peak hours, such as the early morning, AI workloads utilize up to 60% of MIG instances, maximizing infrastructure utilization. The joint training of the LSTM and SAC models ensures smooth transitions between allocation states, thereby mitigating contention during rapid demand shifts.

In the **Poble Sec (PS)** scenario, shown in Figure 8, a residential area with steady demand is modeled. Here, the LSTM model accurately forecasts baseline RAN requirements, allowing the SAC agent to allocate surplus resources ranging from 60% to 85% of MIGs to AI workloads without violating RAN requirement thresholds.

*2) Completion Rate:* The completion rate, which measures the percentage of fulfilled RAN and AI resource demands, underscores the superiority of the proposed CAORA framework over static baselines across all tested scenarios. As illustrated in Figures 9-11, the framework consistently demonstrates near-optimal performance across diverse workload conditions.

The **Les Corts - Camp Nou (LCCM)** scenario, as illustrated in Figure 9, showcases the effectiveness of the CAORA framework in optimizing the utilization of shared resources. The framework successfully meets 98.3% of RAN workload demands while also completing 82% of AI workloads. In contrast, static strategies demonstrate suboptimal resource utilization. The balanced approach only completes 71% of AI workloads, while the RAN-priority strategy fails to properly utilize unused RAN resources, completing just 56% of AI workloads.

Similar patterns are observed in the **El Born (EB)** and **Poble Sec (PS)** scenarios, as shown in Figures 10 and 11. In both cases, CAORA consistently outperforms static strategies in completing both RAN and AI workloads, whereas the RAN-priority and balanced strategies fall short in meeting the required demands. These results validate the effectiveness of the dynamic prioritization mechanism implemented by the proposed SAC agent and highlight the essential role of the LSTM predictor in facilitating proactive resource management.

*3) Adaptability Analysis and Results:* The effectiveness of the proposed SAC-based allocation strategy is quantified using an adaptability metric, which measures its ability to dynamically align resource distribution with real-time demand fluctuations. As shown in Fig. 12, the proposed approach outperforms static baselines across all datasets and scenarios, including Les Corts, Poble Sec, and El Born. This statistically significant advantage is attributed to SAC's ability to learn context-aware policies through continuous interaction with the environment, enabling real-time adjustments based on observed demand states and temporal patterns. In contrast, the fixed allocation in the balanced approach resulted in persistent resource-demand mismatches, while the priority-driven approach was notably inefficient during AI-demand surges, failing to reallocate unused RAN resources. The proposed approach maintained strong performance across varying demand scenarios, particularly excelling during rapid demand transitions where static strategies struggled with delayed responses. This dynamic adaptability is supported by the critic network's learned value function, which estimates the long-term impact of allocation decisions, while the actor network directly maps observed states to optimized resource distributions.

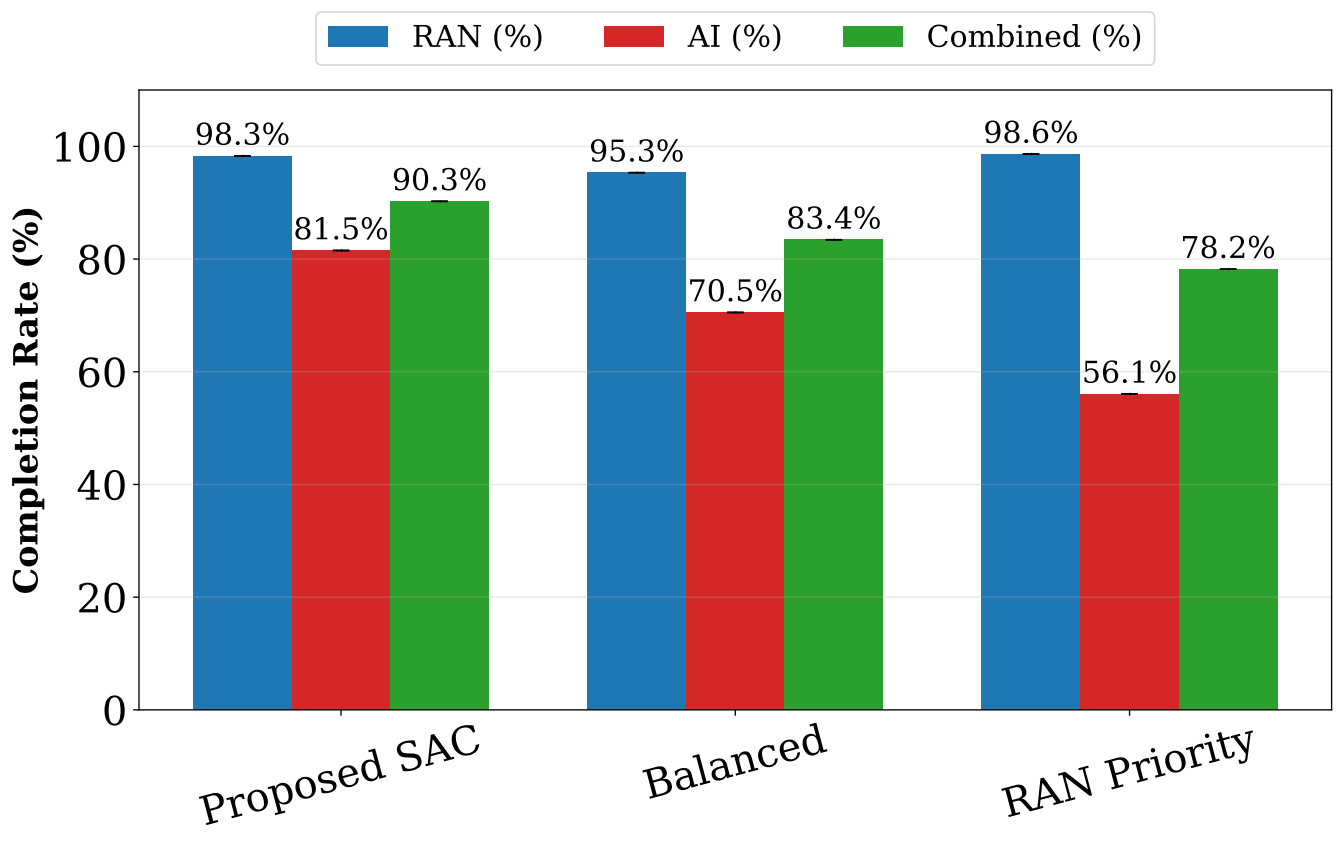

Fig. 9. Completion rate comparison in Les Corts

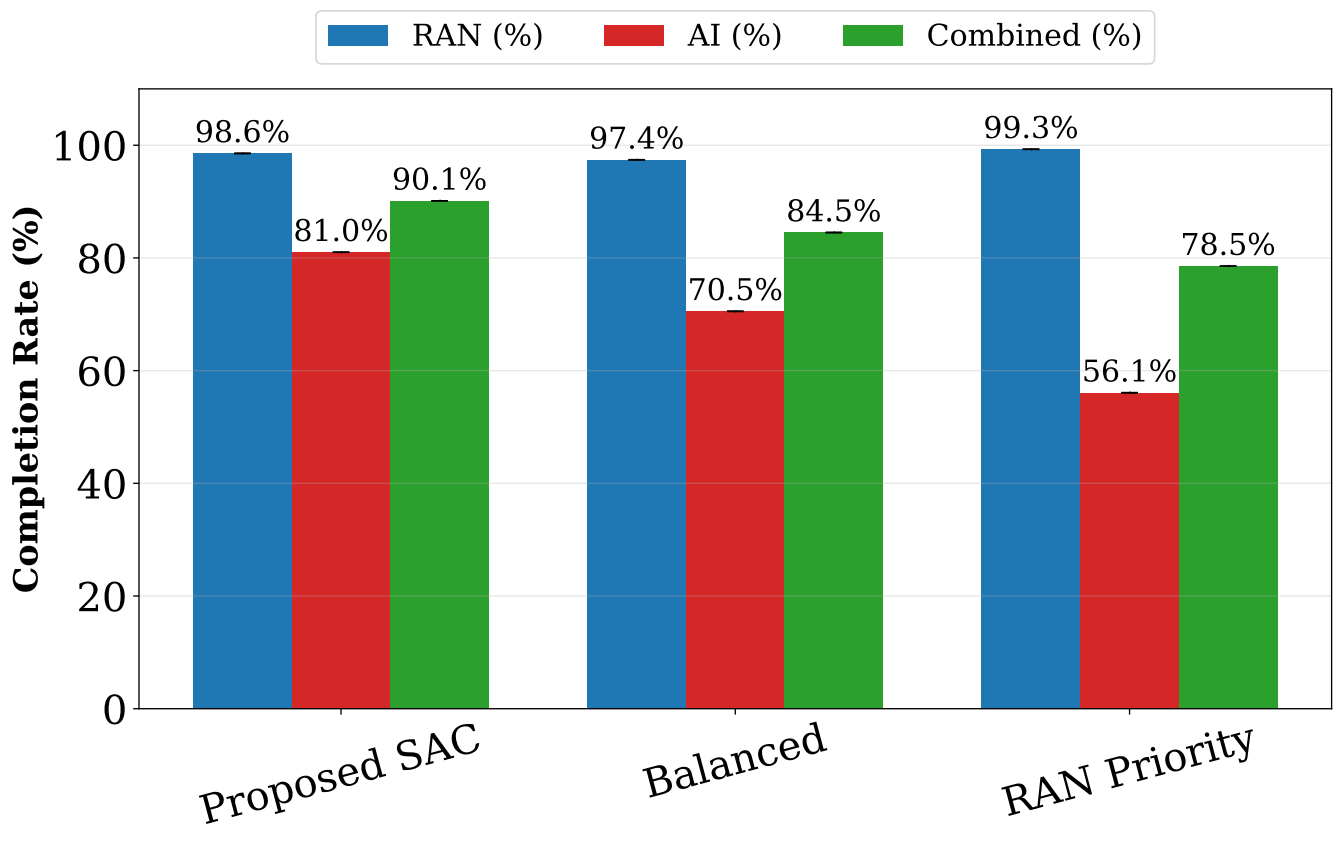

Fig. 10. Completion rate comparison in El Born

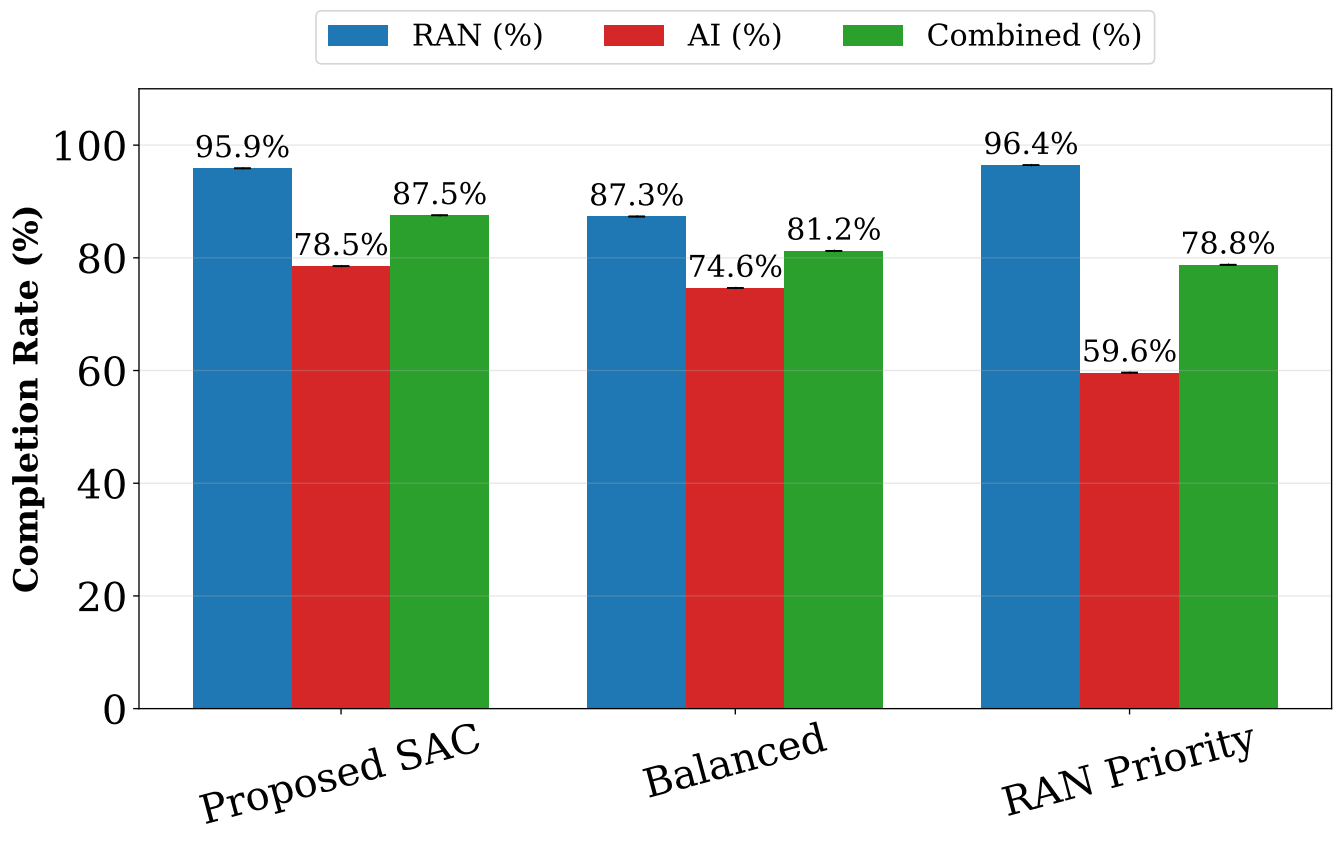

Fig. 11. Completion rate comparison in Poble Sec

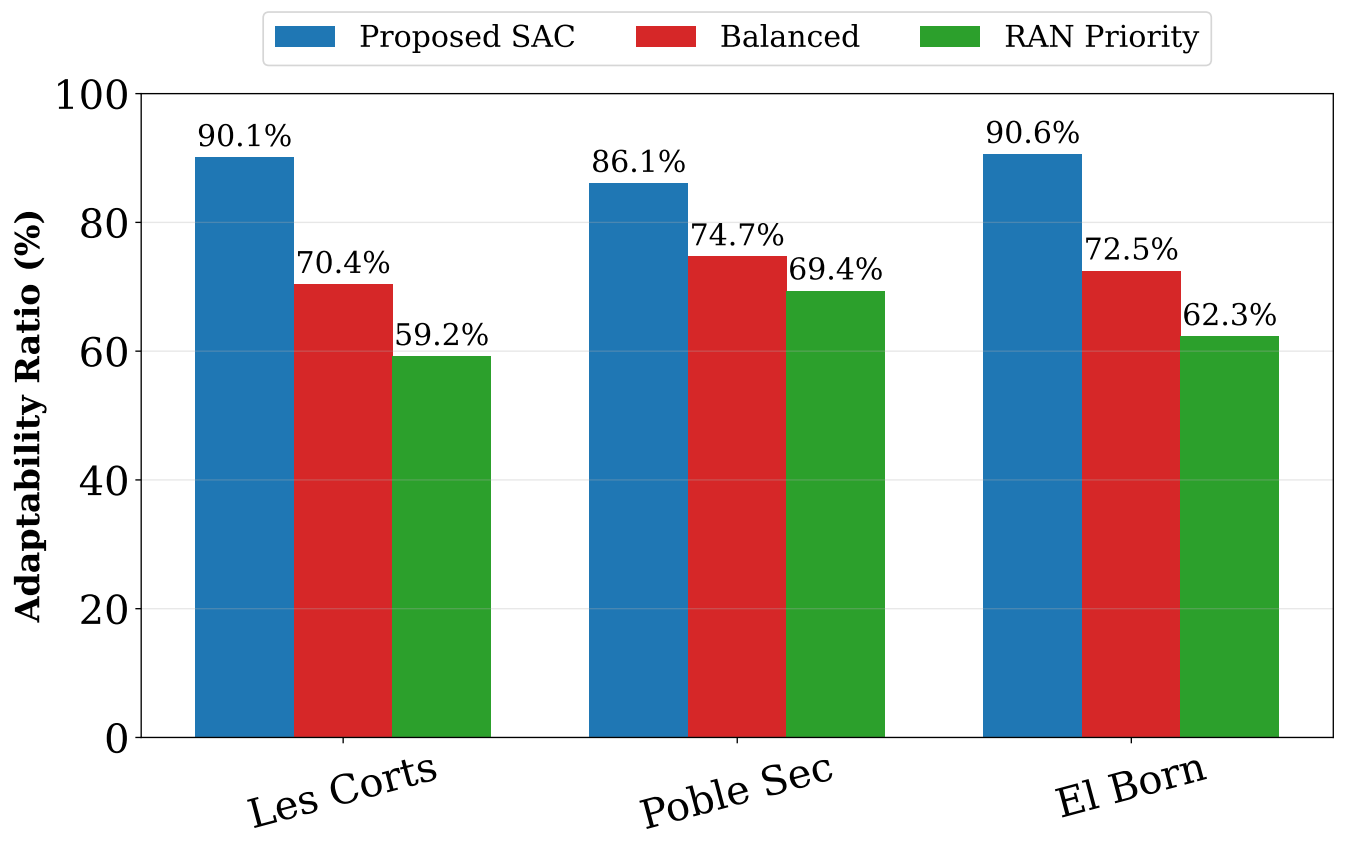

Fig. 12. Performance and adaptability comparison

## VIII. Conclusion and Future Work

In this paper, we presented CAORA, a novel Converged AI-and-ORAN Architectural framework that enables the seamless coexistence of RAN and AI workloads on shared infrastructure. By integrating custom monitoring xApps within the NRT-RIC and leveraging proactive resource orchestration based on LSTM-SAC, CAORA dynamically adapts resource allocation in response to fluctuating network conditions. Our trace-driven evaluations using real-world 5G traffic data confirm that CAORA maintains near-optimal RAN service levels, ensures efficient AI workload support, and maximizes infrastructure utilization even under dynamic and unpredictable load scenarios. These results establish CAORA as a promising architectural framework for future AI-and-RAN converged 6G platforms, providing a foundation for intelligent, efficient, and flexible resource management in next-generation wireless networks.

### A. Future Research Directions

*1) Federated Learning for Scalable and Privacy-Preserving Orchestration:* Extending CAORA to incorporate federated learning could address scalability and data privacy challenges in distributed O-RAN deployments. Collaborative training of LSTM and SAC models across geographically dispersed RAN nodes would reduce centralized data aggregation overhead while accommodating heterogeneous network conditions. Furthermore, integrating federated learning with hierarchical RIC architectures could enable decentralized anomaly detection and resource allocation, enhancing resilience against localized failures or adversarial attacks.

*2) Semantic Communication Integration:* Incorporating AI-driven semantic communication into CAORA could significantly reduce redundant data transmission and improve resource utilization. By embedding semantic relevance metrics, e.g., task criticality, contextual importance, into the SAC reward function, the orchestrator could prioritize resources for high-value AI tasks, such as real-time holographic communications or industrial IoT control. Further, semantic-aware compression could dynamically minimize redundant data flows, freeing resources for latency-sensitive RAN workloads. Integrating intent-based networking interfaces would enable CAORA to interpret high-level service semantics and autonomously translate them into context-aware allocation policies.

## ACKNOWLEDGEMENT

This research was funded by UK Research and Innovation (UKRI), Engineering and Physical Sciences Research Council (EPSRC), under grant number EP/X040518/1 (CHEDDAR project).

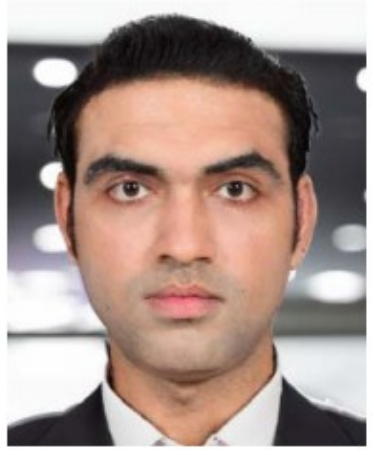

**Syed Danial Ali Shah** He is a Research Fellow at the University of Leeds, U.K., and a Lecturer at Adelaide University, Australia. He earned his M.Eng. from Incheon National University, South Korea, in 2018, and his Ph.D. from RMIT University, Australia, in 2022. He has held positions as a Postdoctoral Research Associate at the University of New South Wales and as a Sessional Academic at RMIT University. His research interests include 5G and beyond wireless networks, IoT, SDN/NFV, vehicular networks, Open RAN, and AI/ML applications in communications. He actively contributes as a reviewer and guest editor for several international journals and conferences.

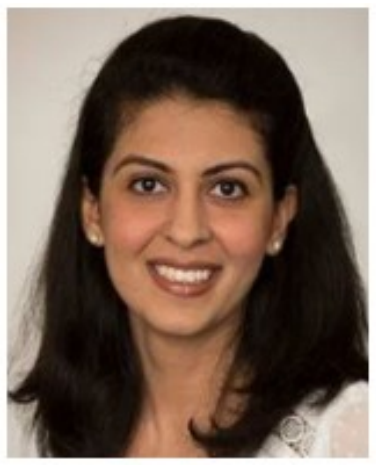

**Maryam Hafeez** She received the Ph.D. degree in Electrical Engineering from the University of Leeds, U.K., in 2015. From 2015 to 2018, she was a Research Fellow with the Institute of Robotics, Autonomous Systems and Sensing, University of Leeds, U.K., where she is currently an Associate Professor of Communication Networks and Systems with the School of Electronic and Electrical Engineering. Her current research is funded by the EU Horizon 2020 programme. Her research interests include the design and analysis of protocols for next-generation green intelligent wireless networks, applying tools from game theory and stochastic geometry, as well as research related to the Internet of Things and Industry 4.0. She currently serves on the Editorial Board of Frontiers in Communications and Networks.

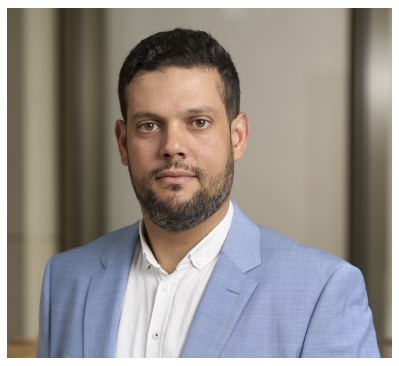

**Abdelaziz Salama** He received the B.Sc. degree in Electrical and Electronic Engineering from Tripoli University, Libya, in 2009, and the M.Sc. degree in Communication, Control, and Digital Signal Processing from the University of Strathclyde, Glasgow, U.K., in 2017. He completed his Ph.D. at the University of Leeds, U.K., in 2024, where he is currently a Research Fellow. His research interests include federated learning, autonomous systems, and sensing. He also has nine years of industry experience in telecommunications engineering, information technology, and management across local and international firms.

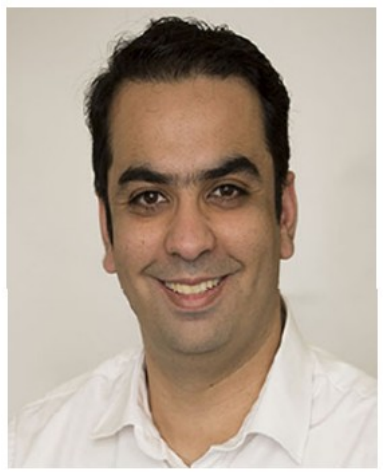

**Syed Ali Raza Zaidi** He received his Doctoral degree from the School of Electronic and Electrical Engineering and is currently an Associate Professor at the University of Leeds, working in the areas of communications and sensing for robotics and autonomous systems (RAS). From 2011 to 2015, he held research positions with the International University of Rabat, the SPCOM Research Group (US ARL-funded), and Qatar Innovations and Mobility Centre (QNRF-funded). He has authored over 90 papers in leading IEEE journals and conferences. His research interests span ICT, applied mathematics, mobile computing, and embedded systems. He is a recipient of the G. W. and F. W. Carter Prizes and has secured funding from COST, Royal Academy of Engineering, EPSRC, Horizon EU, and DAAD. His editorial experience includes IEEE Communication Letters, IET Signal Processing, IET Access, and IEEE Communication Magazine, where he currently serves as Associate Technical Editor.